\documentclass[twocolumn,showpacs,aps,amsmath,amssymb,prb]{revtex4}

\usepackage{graphicx,epsf}
\usepackage{bm}

\begin{document}


\title{Current--phase relation in Josephson junction coupled with a magnetic dot}

\author{A.~V.~Samokhvalov}

\affiliation{Institute for Physics of Microstructures,
Russian Academy of Sciences,
603950, Nizhny Novgorod, GSP-105, Russia}

\date{\today}                   
\begin{abstract}
The current-phase relation for a short Josephson junction placed
in the nonuniform field of a small ferromagnetic particle is studied.
The effect of the particle produced on the junction appears
to be strong due to the formation of the pair of oppositely
directed Abrikosov vortices which pierce the
thin film superconducting electrode and cause
a small--scale inhomogeneity of Josephson phase difference.
The induced phase difference inhomogeneity is shown
to result in the nonzero fixed phase drop
$\varphi_0$ across the junction.
The equilibrium value $\varphi_0$ corresponding
to the ground state of the junction depends the configuration
of the vortex--antivortex pair.
The possibility to tune the ground state phase difference
$\varphi_0$ is discussed.
\end{abstract}

\pacs{Valid PACS appear here}   
\maketitle



\section{\label{sec:Intro}{Introduction}}

Usually the current--phase relation (CPR) in
Josephson junction
close to  the critical temperature is sinusoidal
$I_s(\varphi) = I_c \sin\varphi$,
and the dependence of the free energy
$E_J = (\hbar I_c / 2 e) ( 1 - \cos\varphi )$
assumes positive values of the critical current $I_c > 0$
(see Ref.~\onlinecite{Golubov-RMP04}). %
So, in the absence of a supercurrent, $I_s = 0$,
the phase drop across the conventional junction equals zero
\cite{Josephson-PL62}. %
But under certain conditions one can fabricate so-called Josephson $\pi$ junction
\cite{Bulaevskii-JETPL77}
which has an energy minimum at $\varphi=\pi$,
i.e., it provides a phase shift of $\pi$
in the ground state
(Refs.~\onlinecite{Golubov-RMP04,Buzdin-RMP05}).
The CPR of $\pi$ junctions reads
$I_s(\varphi) = I _c\,\sin(\varphi+\pi)$
and can be formally described by the negative value of the
critical current $I_c$.
The $\pi$ states have been observed in 
Josephson junctions consisting of two $d-$wave superconductors
\cite{Tsuei:00} 
in superconductor/ferromagnet/superconductor (SFS) junctions
utilizing ferromagnetic barriers
\cite{Ryazanov-PRL01,Kontos-PRL02,Blum-PRL02},
and also in multiterminal superconductor/normal-metal/superconductor (SNS)
Josephson junctions 
\cite{Baselmans-Nature00}.
Such $\pi$ junctions are supposed to open up new opportunities
for designing Josephson effect--based devices
\cite{Ioffe-Nature99,Yamashita-PRL05,Ustinov-JAP03}. %

Recently the investigations of Josephson $\varphi$ junctions
which provide the realization of an unusual current--phase relation
\begin{equation}\label{eq:i}
    I_s(\varphi) = I_c\,\sin(\varphi+\varphi_0)
\end{equation}
have been attracting  a lot of attention
\cite{Buzdin-PRB03,Goldobin-PRB07,Buzdin-PRL08}.
The minimum of the Josephson energy of $\varphi$ junctions
$E_J = (\hbar I_c / 2 e) [ 1 - \cos(\varphi+\varphi_0) ]$
corresponds to the nonzero value of the phase
difference $\varphi = -\varphi_0$ such as
$0 < \varphi_0 < \pi$.
The realization of the $\varphi$ junction is possible
in the case of periodic structures composed of alternating
$\mathrm{0}$ and $\pi$ mini--junctions
\cite{Buzdin-PRB03,Goldobin-PRB07},
or in the case of SNS structures when
the normal layer is a noncentrosymmetric
magnetic metal
\cite{Buzdin-PRL08}.
Josephson $\varphi$ junctions demonstrate unusual properties
and may serve as phase shifters
in the superconducting (SC) electronics circuits
\cite{Terzioglu-ASC98}.

The $\pi(\varphi)$ junctions described above utilize
an intrinsic phase change due to the peculiarities
of tunneling through the ferromagnetic layer
or/and superconducting wavefunction symmetry.
An alternative approach is to produce a phase shift
across the junction using flux- or current--biasing.
Examples are tools based on trapping fluxoids in
a mesoscopic ring incorporated into dc-SQUID
\cite{Majer-APL02}
or Josephson junction with the additional current
injector--extractor pair
which creates an arbitrary discontinuity
of the Josephson phase difference
\cite{Ustinov-APL02,Goldobin-PRL04}.
Here we suggest to use small ferromagnetic particles
to create the phase--biased Josephson system.
An arbitrary phase drop across the junction
is shown to be caused by a small--scale phase
difference inhomogeneity induced by the particle.

Let us briefly remind of the basic mechanisms which could provide a strong phase variation along the contact on a scale
which is smaller than the Josephson penetration depth.
First of all, the natural source of a phase inhomogeneity is an
Abrikosov vortex (AV) pinned in the SC electrodes.
Even a single misaligned AV, trapped in the SC
electrodes perpendicular to the junction plane,
is known to modify strongly the critical current and the
current--voltage characteristics of Josephson junctions
\cite{Uchida-JAP83,Miller-PRB85,Golubov-JETP87,Ustinov-PRB93,Fistul-PRB98}.
Next, $0-\pi$ discontinuities in the phase difference
along the barrier appear in a Josephson junction
composed of alternating $0$ and $\pi$ mini-junctions
\cite{Mints-PRB97} %
(the zigzag junctions between high-$T_c$ and conventional superconductors
\cite{Smilde-PRL02} %
as well as SFS junctions with a steplike thickness
of the ferromagnetic interlayer
\cite{Goldobin-PRL06}). %
Finally, arbitrary phase inhomogeneity can be
formed with a current injection into Josephson junction
on a scale smaller than the characteristic Josephson length
\cite{Ustinov-APL02,Goldobin-PRL04}.
The presence of such phase singularities results in
the unusual CPR 
\cite{Buzdin-PRB03}, %
an anomalous non-Fraunhofer $I_c(H)$ dependence
\cite{Smilde-PRL02}, %
and spontaneous generation of fractional Josephson vortices
at the boundaries between $0$ and $\pi$ regions %
\cite{Mints-PRB98,Mints-PRL02}.

Another method for creating of a controlled
phase inhomogeneity in Josephson junctions
has been proposed and successfully realized recently.
This method is based on the interaction of a Josephson contact
with the nonuniform magnetic field of submicron ferromagnetic
particles located close to the junction
\cite{Aladyshkin-JMMM03,Vdovichev-JETPL04,Fraerman-PRB06}.
These experiments have demonstrated a essential dependence
of the critical current $I_c$ on the
magnetic state of the particles.
This means that the transport properties of Josephson contacts can be effectively controlled by the magnetic field
of the small particles.
Experimentally detected additional maxima
in the field dependence of the
critical current $I_c(H)$ have unambiguously indicated
commensurability effects between a periodic
distribution of the Josephson phase difference
created by the particles and the scale of the
phase modulation induced by an applied magnetic field $H$
\cite{Fraerman-PRB06,Samokhvalov-JETP07}.
While the macroscopic commensurability effects have already been
demonstrated in such hybrid ferromagnet--superconductor (FS) systems,
the insight into the current--phase relation is still lacking. %

In this paper we study theoretically the hybrid FS system
consisting of a Josephson junction coupled with a single
magnetic dot as it is shown in Fig.\ref{Fig:1}. %
We demonstrate that the phase shift in the ground state of the
Josephson junction depends on the magnetic state of the particle.
We discuss the possibility to realize of Josephson $\varphi$-junction
based on such hybrid FS structure.
The paper is organized as follows.
In Sec.\ref{sec:Model}, we introduce a model used to explain the appearance
of a nonuniform phase--difference distribution in Josephson junction coupled with a magnetic dot.
In Sec.\ref{sec:Phase}, we study the ground state of this hybrid system
which is characterized by the finite phase difference
drop  $\varphi_0$ across the junction.
In Section \ref{sec:Concl} we summarize our results.

\section{\label{sec:Model}{Josephson phase modulation induced by a
magnetic particle}}

We consider a generic example of the FS hybrid system consisting
of a short square ($W\times W$)
Josephson junction and an elongated magnetic particle
on its top electrode.
The junction is formed by overlapping two long SC strips
($L \gg W$) of thickness $d \ll \lambda$ (top) and
$D \gtrsim 2\lambda$ (bottom) as shown in Fig.\ref{Fig:1}a.
The single-domain magnetic dot with in-plane magnetization
$\mathbf{M}$ is separated from the top SC electrode
by a thin insulating layer, which prevents the proximity effect.
The interaction between the junction and the particle
may be provided by the magnetic field generated by
the dot and supercurrents.
For the sake of simplicity, we consider here only the
case of a rather small junction with
\begin{equation}\label{eq:a1}
    \lambda \ll W \ll \Lambda,\,\lambda_J\,,
\end{equation}
where $\lambda$ and $\lambda_J$ are the London penetration depth
and the Josephson penetration depth, respectively,
and $\Lambda=\lambda^2/d$ is the thin--film screening length.
The gauge invariant phase difference across the junction
is given by the expression
\begin{equation}\label{eq:a2}
    \phi(\mathbf{r}) = \theta^b(\mathbf{r}) - \theta^t(\mathbf{r})
    + \frac{2\pi}{\Phi_0} \int\limits_b^t\, dz\, A_z(\mathbf{r})\,.
\end{equation}
Here $\theta^{b}(\mathbf{r})$ and $\theta^{t}(\mathbf{r})$
are distributions of the phase of Cooper wave function
in the bottom and the top SC electrodes, respectively,
$\mathbf{r} = (x, y)$ is a vector in the junction plane,
$A_z$ is the normal to the junction plane
component of the magnetic vector potential
$\mathbf{A}=\mathbf{A}_\parallel + A_z \mathbf{z}_0$, and
$\Phi_0=\pi \hbar c /\,|\,e\,|$ -  is the flux quantum.
In a thin SC strip with $W \ll \Lambda$, the self--field of the
sheet current can be disregarded, and the top electrode
is assumed to be transparent to the magnetic field of the particle.
In its turn, this magnetic field partially penetrates
into the bottom superconductor and induces in-plane
screening Meissner currents in it.
Since the top film thickness $d$ is assumed to be small
it is reasonable to apply the gauge $A_z = 0$ to avoid the strong singularity
of the vector potential $\mathbf{A}$ in the limit $d\to 0$.
For this gauge the phase difference across the junction
$\phi(\mathbf{r})$ (\ref{eq:a2})
is determined by the distributions
$\theta^{t}(\mathbf{r})$ and $\theta^{b}(\mathbf{r})$
of the phase of the wave functions only.

In the absence of vortex lines trapped in the electrodes
the ground state of the system can be described
by the uniform phase
$\theta^t = \theta^b = 0$
and screening Meissner currents are determined by the
the in-plane component of the vector potential
$\mathbf{A}_\parallel$.
So, the gauge invariant phase difference (\ref{eq:a2})
across the junction equals zero:
$\phi(\mathbf{r})=0$,
and the magnetic field of the particle does not modify
the critical current of the Josephson junction.

Vortex (antivortex), if appears, must be located
near the negative (positive) pole of the magnetic dot
\cite{VanBael-PRL01},
as shown in Fig.~1a.
According to the concept proposed in %
Ref.\onlinecite{Fraerman-PRB06},
a pair of vortices of opposite directions pierce
the top electrode of the junction.
As before, we put $\theta^b = 0$ due to the absence
of vortex lines in the bottom electrode.
Suppose that the pair size (i.e., the vortex--antivortex distance)
$\vert \mathbf{r}_v - \mathbf{r}_a \vert = 2a$
is rather large compared with the superconducting coherence
length $\xi$, then the electrodynamic mechanism
based on the spatial dependence of the gauge
invariant phase difference
\begin{equation}\label{eq:a4}
    \phi(\mathbf{r}) = -\theta^t(\mathbf{r})
\end{equation}
is dominant
\cite{Miller-PRB85,Golubov-JETP87}.
Here $\mathbf{r}_v=(x_v,\,y_v)$ and $\mathbf{r}_a=(x_a,\,y_a)$
are the vortex and antivortex positions, respectively.

\subsection{\label{sec:Approach}{Basic equations}}

As a next step we should find the phase difference distribution
$\phi(\mathbf{r})$
over the junction area.
The small size of the junction $W \ll \lambda_J$ means
that self--field effects of the Josephson current
can be neglected compared to the in-plane currents.
In this case, the phase-difference distribution $\phi(\mathbf{r})$ obeys
the two-dimensional Laplace equation
\cite{Barone:Joseph,Barone-JAP82}:
\begin{equation}\label{eq:1}
    \triangle\phi = \partial_x^2\phi + \partial_y^2\phi = 0.
\end{equation}
In the presence of trapped Abrikosov
vortices the top electrode of the Josephson contact becomes a
multiply connected domain.
So, it is necessary to take into account the topological
singularities of the phase distribution $\theta^t(\mathbf{r})$
which are caused by presence of the vortex--antivortex pair
\cite{Bulaevskii-PRB92}.
According to the Eq.~(\ref{eq:a4})
we obtain the following condition for $\phi(\mathbf{r})$:
\begin{equation}\label{eq:a3}
    {\rm curl}_z(\nabla\phi) = 2\pi
       \left[\,\delta(\mathbf{r}-\mathbf{r}_a)%
             -\delta(\mathbf{r}-\mathbf{r}_v)\, %
       \right]\,,
\end{equation}
which fixes the circulation around the singularities.

For $W \ll \Lambda$, the in-plane sheet current density in the top
electrode
\begin{equation}\label{eq:a5}
    \mathbf{\sigma}^t = -\frac{c \Phi_0}{8 \pi^2 \Lambda}
        \left(\nabla\theta^t + \frac{2\pi}{\Phi_0} \mathbf{A}_{\|}\right)
\end{equation}
is determined mainly by the phase gradient $\nabla\theta^t$
induced by the trapped vortices
rather than the vector potential $\mathbf{A}_{\|}$.
Indeed, the term with $\mathbf{A}_{\|}$
in the Eq.~(\ref{eq:a5}) is of the order of
$2\pi / \Lambda \ll \vert \nabla\theta^{\,t} \vert \sim 2\pi / W$
and can be neglected.
It means that the sheet current density (\ref{eq:a5})
is determined by the gauge invariant phase difference $\phi(\mathbf{r})$:
$\mathbf{\sigma}^t \sim \nabla\,\theta^t \sim \nabla\phi$.
At the edges of the top stripe ($x=0,W$\,; $y=0$)
the normal component of the sheet current
$\mathbf{\sigma}_t$ vanishes, and the Eq.~(\ref{eq:1})
must be supplemented with the following boundary conditions:
\begin{equation} \label{eq:2}
    \partial_x \phi \bigg\vert_{x=0,W} = 0, \quad %
    \partial_y \phi \bigg\vert_{y=0} = 0.
\end{equation}
Finally, a local phase inhomogeneity due to the presence
of a vortex--antivortex pair has to vanish
at distances larger as compared with the pair size $2a$.
For the sake of simplicity the top electrode is assumed to be
a semi--infinite SC strip, and the condition
\begin{equation} \label{eq:3}
    \phi(x, y) = 0\,,\quad \mathrm{for}\:\: y \to \infty\,
\end{equation}
has to be satisfied.
Thus, the equations (\ref{eq:1}),(\ref{eq:a3}) and the boundary
conditions (\ref{eq:2}),(\ref{eq:3}) describe the phase
difference distribution which is induced by a vortex--antivortex pair
trapped in the top electrode of the contact.

At a constant value of the critical current density $j_c$
and the standard sinusoidal form of CPR
the ground state of this junction in the absence of a supercurrent
corresponds to a minimum of the Josephson energy
\begin{equation}\label{eq:6}
    E_J(\varphi) = \frac{\hbar I_c}{2 e}
        - \frac{\hbar j_c}{2 e} \int\limits_{S_J} d\mathbf{r}\,
       \cos\left(\varphi+\phi(\mathbf{r})\right)\,,
\end{equation}
where the integral is evaluated over the junction area
$S_J:(0 \le x,y \le W)$, and $I_c = j_c\, S_J$.
The current--phase relation for the junction
\begin{equation}\label{eq:5}
    I_J(\varphi) = j_c \int\limits_{S_J} d\mathbf{r}\, %
       \sin\left(\varphi+\phi(\mathbf{r})\right),
\end{equation}
reveals the shift which depends on the phase difference
distribution $\phi(\mathbf{r})$.

By minimizing the Josephson energy (\ref{eq:6})
with respect to $\varphi$
one can find the equilibrium distribution
$\phi_e(\mathbf{r}) = \phi(\mathbf{r}) + \varphi_0$
of the gauge invariant phase difference for the junction
containing the vortex--antivortex pair,
where the additive constant
\begin{equation}\label{eq:12}
    \varphi_0 = - \mathrm{arctan}\left(S_\phi /  C_\phi\right)\,,
\end{equation}
determines the fixed phase shift 
between the top and bottom electrodes away from the junction area.
The coefficients $C_\phi$ and $S_\phi$ in (\ref{eq:12})
depend on the distribution of the phase difference
$\phi(\mathbf{r})$ induced by the trapped Abrikosov vortices:
\begin{equation}\label{eq:13}
    C_\phi = \int\limits_{S_J} d\mathbf{r}\, \cos\phi(\mathbf{r})\,,\quad
    S_\phi = \int\limits_{S_J} d\mathbf{r}\, \sin\phi(\mathbf{r})\,.
\end{equation}
Varying the magnetic state of the particle one may control
$\phi(\mathbf{r})$ changing the vortex(antivortex) position, and,
thereby, tune the average phase difference $\varphi_0$
across the junction.

\subsection{\label{sec:Model1}{Phase Difference Distribution}}

The the gauge invariant phase difference $\phi(\mathbf{r})$
due to the presence of a single vortex--antivortex pair
trapped in the top electrode can be calculated as follows.
First of all, let's neglect the edge effects (\ref{eq:2}).
The distribution $\phi(\mathbf{r})$, which satisfies
the Laplace equation (\ref{eq:1}) and provides for
the required phase circulation (\ref{eq:a3})
around the singularities $\mathbf{r}_{v,a}$,
can be written as a superposition
of contributions from two point opposite vortices:
\begin{equation}\label{eq:a7}
     \phi(\mathbf{r}) = \phi_p(\mathbf{r})\,,\quad
     \phi_p(\mathbf{r}) = \theta_a(\mathbf{r}) - \theta_v(\mathbf{r})\,.
\end{equation}
The phase $\theta_{v,a}(\mathbf{r})$ of the wave function
describing a point vortex (antivortex)
is determined by the polar angle
specifying the direction from the position of the vortex axis
$\mathbf{r}_{v,a}$ to the reference point $\mathbf{r}$
(see Fig.~\ref{Fig:1}b):
\begin{equation}\label{eq:7}
    \theta_{v,\,a}(\mathbf{r}) = \mathrm{arctan}\left( %
    \frac{y-y_{v,\,a}}{x-x_{v,\,a}} \right)\,.
\end{equation}
It is evident that $\phi_p(\mathbf{r}) \to 0$ for
$| \mathbf{r} | \gg | \mathbf{r}_{v,a} |$ and, thus,
the distribution (\ref{eq:a7}),(\ref{eq:7}) satisfies the condition
(\ref{eq:3}).
Figure~\ref{Fig:2} illustrates schematically the distribution
of the phase difference $\phi_p(\mathbf{r})$ created by a pair
of opposite vortices.
The dark central area between the vortices shows the
region where $\pi/2 < \phi_p < 3\pi/2$ and $\cos\phi_p < 0$.
This domain provides with an additional
positive contribution to the Josephson energy (\ref{eq:6})
and, as a consequence, such state of the junction appears
to be an energetically unfavourable.
The energy excess associated with the pair presence
grows as the intervortex distance
$\vert \mathbf{r}_v - \mathbf{r}_a \vert$ increases.

To take into account the boundary conditions
(\ref{eq:2}) the solution $\phi(\mathbf{r})$
may be written in the following convenient form:
\begin{equation}\label{eq:8}
    \phi(\mathbf{r}) = \phi_p(\mathbf{r})
                      + \phi_p^\prime(\mathbf{r})
                      + \psi(\mathbf{r})\,.
\end{equation}
Here
$\phi_p^\prime(\mathbf{r}) = \theta_v^\prime(\mathbf{r})
-\theta_a^\prime(\mathbf{r})$
is the phase distribution created by the vortice images
(see Fig.~\ref{Fig:1}b)
\begin{equation}\label{eq:a8}
    \theta_{v,\,a}^\prime(\mathbf{r})= \mathrm{arctan}\left( %
    \frac{y+y_{v,\,a}}{x-x_{v,\,a}} \right)
\end{equation}
and $\psi(\mathbf{r})$ is the solution of the Laplace equation
\begin{equation}\label{eq:9}
    \triangle\,\psi = 0
\end{equation}
in the infinite stripe
( $0 \le x \le W$, $\vert\, y\, \vert < \infty$ )
with the following boundary conditions at the stripe edges
$x=0$ and $x=W$
\begin{equation}\label{eq:10}
    \partial_x \psi \bigg\vert_{x=0,W}
    = - \partial_x\,(\phi_p + \phi_p^\prime)\bigg\vert_{x=0,W}\,:
\end{equation}
%
%
%
\begin{widetext}
\begin{eqnarray}\label{eq:11}
    \psi(x,\,y)&=&\frac{1}{\pi}\int\limits_{-\infty}^{+\infty} du\,
    \left[ \frac{y_v (u^2-r_v^2)}{(r_v^2+u^2)^2-4y_v^2u^2}\:
          -\frac{y_a (u^2-r_a^2)}{(r_a^2+u^2)^2-4y_a^2u^2} \right]
          \ln\left[ \cosh\left(\pi\frac{y-u}{W}\right)
                   - \cos(\pi\frac{x}{W}) \right] \\
               &-&\frac{1}{\pi}\int\limits_{-\infty}^{+\infty} du\;
    \left[ \frac{y_v (u^2-p_v^2)}{(p_v^2+u^2)^2-4y_v^2u^2}\:
          -\frac{y_a (u^2-p_a^2)}{(p_a^2+u^2)^2-4y_a^2u^2} \right]
          \ln\left[ \cosh\left(\pi\frac{y-u}{W}\right)
                   + \cos(\pi\frac{x}{W}) \right]\,,\nonumber
\end{eqnarray}
\end{widetext}
where
$$
    r_{v,a}^2 = x_{v,a}^2 + y_{v,a}^2\,, \qquad
    p_{v,a}^2 = (W-x_{v,a})^2 + y_{v,a}^2\,.
$$

The expressions
(\ref{eq:a7})-(\ref{eq:a8}),(\ref{eq:11}) determine
the distribution of the gauge invariant phase
difference $\phi(\mathbf{r})$
created by the pair of opposite point vortices
trapped in the thin top electrode of the Josephson junction.
This phase distribution $\phi(\mathbf{r})$
is used to calculate the Josephson energy (\ref{eq:6})
and the total current through
the contact (\ref{eq:5})
for different positions of vortex and antivortex.


\section{\label{sec:Phase}{Ground state of Josephson
junction coupled with a magnetic particle}}

Now we proceed with the calculations of the ground state
of the Josephson junction which depends on the
size and orientation of the vortex--antivortex pair
induced by the magnetic particle.
We restrict ourselves to the case of zero
homogeneous external magnetic field.
For simplicity, the vortex and antivortex
are assumed to be placed symmetrically
with respect to the center of the junction
$x_0=y_0=W/2$:
$$
    x_v + x_a = W, \quad y_v + y_a = W\,.
$$

Figures~\ref{Fig:3} shows the dependence of the
average phase difference $\varphi_0$ (\ref{eq:12})
and the Josephson energy $E_J(\varphi_0)$ (\ref{eq:6})
on the location of the vortices.
As an example we consider the change in the phase $\varphi_0$
due to the rotation of the vortex--antivortex pair
with respect to the midpoint $x_0=y_0=W/2$.
In this case the location of the vortices is determined
by the intervortex distance $2a$ and the angle of the pair rotation
$$
    \alpha=\mathrm{arctan}\left(\frac{y_v-y_a}{x_v-x_a}\right)
$$
relative to the direction of $x$-axis (see Fig.~\ref{Fig:1}).
The range of the change in $\varphi_0$ depends on the pair size $2a$.
For $2a \ll W$ (Fig.~\ref{Fig:3}, curve 1) the phase
inhomogeneity occupies a small part of the junction area,
and the presence of the vortex--antivortex pair does not
affect the junction properties essentially.
The value of $\varphi_0$ varies weakly round about the point
$\varphi_0 = 0$, and the junction demonstrates mainly the
properties of a conventional junction:
the ground state corresponds to the almost zero phase drop
across the junction.
An increase in the pair size leads to
forming a strong phase modulation $\phi(\mathbf{r})$, and
the average phase difference $\varphi_0$ can take
practically any value between $-\pi/2$ and $\pi/2$
in dependence on the angle $\alpha$
(Fig.~\ref{Fig:3}, curve 2).
This is accompanied by the growth of the Josephson energy
$E_J$ due to the expansion of the domain where
$\cos\phi(\mathbf{r}) < 0$.
The further increase in the pair size ($2a > W/2$) leads
to the additional $\pi$ shift of the average phase
difference $\varphi_0$ (Fig.~\ref{Fig:3}, curve 3),
and the junction exhibits the switching into
the new ground state which is specific for $\pi$ junctions:
the value of $\varphi_0$ oscillates about the point
$\varphi_0 = \pi$.

Figure~\ref{Fig:4} shows the dependences of the average phase
difference $\varphi_0$ and the Josephson energy of the ground state
$E_J^{\varphi_0}$ vs the intervortex distance $2a$
for the fixed orientation
of the vortex-antivortex pair $\alpha=\pi/2$.
For a small intervortex distance $2a$, the domain, where
the energetically unfavourable phase difference
$\pi/2 < \phi(\mathbf{r}) < 3\pi/2$ exists,
occupies the central part of the junction.
The size of the domain is small,
and this part of the junction
gives a small additional positive contribution
to the Josephson energy $E_J$. %
As a result, the conventional ground state realizes,
which is described by the zero value of the average phase
difference $\varphi_0$.
The increase in the intervortex distance $2a$ provokes
an expansion of the energetically unfavourable domain,
and, consequently, the rise in the Josephson energy
$E_J^0$ of the state corresponding to a choice of
$\varphi_0=0$.
Let's introduce the complementary phase difference distribution
$\phi^*(\mathbf{r})$ which differs from
the initial one $\phi(\mathbf{r})$ by the $\pi$ shift:
$\phi^*(\mathbf{r})=\phi(\mathbf{r})+\pi$.
In the absence of trapped vortices the complementary state
is unrealizable, since it corresponds to a maximum
of the Josephson energy (\ref{eq:6}).
The generation of the vortex-antivortex
pair results in a decrease of the Josephson energy $E_J^{\pi}$
of the complementary state $\phi^*(\mathbf{r})$.
The crossing of the curves $E_J^0(2a)$ and $E_J^{\pi}(2a)$
occurs at $2a = 2a^* \simeq 0.56\,W$, so that the switch between
the conventional state ($\varphi_0 = 0$) and the $\pi$ shifted
one ($\varphi_0 = \pi$) takes place.
As a result, the ground state of the junction changes
for sufficiently large intervortex distances $2a > 2a^*$.
Certainly, the switching point $2a^*$ depends on configuration
of the vortex-antivortex pair.
Such a crossover between $0$ and $\pi$ states
manifests itself as a $\pi$ shift of the
superconducting phase between the electrodes
of the junction.

In Fig.~\ref{Fig:5} we present examples of the simulations
of the equilibrium phase difference
$\phi(\mathbf{r})+\varphi_0$
for the junction containing the vortex--antivortex pair.
For a small intervortex distance ($2a < 2a^*$),
the conventional ground state $\varphi_0 = 0$
realizes (Fig.~\ref{Fig:5}a).
This state corresponds to the zero phase drop across the junction
in the absence of a supercurrent.
With the increase in the intervortex distance ($2a > 2a^*$),
the additive phase shift $\varphi_0 = \pi$ decreases
in part the Josephson energy corresponding the central domain
between the vortices (Fig.~\ref{Fig:5}b).
As a result, there appears a $\pi$ shift in the phase
of the superconducting order parameter
across the junction.

\section{Summary} \label{sec:Concl}

In conclusion, we have studied theoretically the properties
of the hybrid system consisting of a short Josephson junction
located in a nonuniform field of the magnetic particle.
The effect of the particle on the junction is shown
to be strong due to the formation of the pair of oppositely
directed Abrikosov vortices which pierce the top electrode
of the junction.
From an experimental point of view the vortex--antivortex pair
can be created by cooling the junction through transition
temperature $T_c$ in the dipole magnetic field of a magnetic dot
\cite{VanBael-PRL01}.
As a result, one should use single--domain magnetic dots
with in--plain magnetization such as elongated submicron
$\mathrm{Co}$ dots.
For parameters of magnetic dots taken from the experiments,
\cite{Fraerman-PRB06}
(the saturation magnetization
$M_s \sim 800\, {\rm Oe}$,
lateral dimensions
$\sim 650\,(\mathrm{easy\; axis}) \times 250\, \mathrm{nm^2}$
and a thickness of
$\sim 50\, \mathrm{nm}$)
we can estimate typical dimension of the Josephson junction as
$W \sim 1\, \mathrm{\mu m}$.
Magnetostatic calculations show that the stray field of both
poles of the dot creates a (positive or negative) flux
$\Phi_s > \Phi_0$ through the top SC electrode,
and the criterion
\cite{VanBael-PRL01,Milosevic-PRB04}
for the nucleation of the
vortex--antivortex pair
is satisfied.
Since the flux $\Phi_s$ rapidly decreases
when a distance between the magnetic dot and a superconductor
becomes comparable with the lateral sizes of the dot,
the Meissner state is more energetically favorable
in the thick bottom electrode upon cooling through $T_c$.

The vortex--antivortex pair trapped in the top electrode
causes the inhomogeneity
of the gauge invariant phase--difference on scales that
are significantly smaller than the Josephson length.
The pair size and orientation are determined by the size
and magnetization of the particle.
We have calculated the corresponding distribution
of the Josephson phase difference over the junction
and have studied the dependence of the Josephson energy
$E_J$ on the vortex configuration.
It was shown that the ground state of the junction
corresponds to the nonzero drop of the average phase
difference across the junction $\varphi_0$.
The equilibrium value of the phase difference $\varphi_0$
depends on the size and orientation of the vortex--antivortex
pair, i.e. on the magnetic state of the particle.
This demonstrates the possibility to realize the
tunable current--phase relation
$I_s=I_c \sin(\varphi+\varphi_0)$
by changing the magnetization of the particle.
Finally, note that such hybrid FS system incorporated
into a superconducting ring should induce spontaneous
currents and may serve as a natural phase shifter
in the superconducting electronic circuits.

\begin{acknowledgments}
I would like to thank  A.\ S.\ Mel'nikov for
critical reading of the manuscript
and for valuable suggestions.
I am thankful to A.\ I.\ Buzdin and A.\ A.\ Fraerman
for stimulating discussions.
This work was supported, in part, by the Russian Foundation
for Basic Research, and by the program "Quantum Physics of Condensed Matter" of the Russian Academy of Sciences.

\end{acknowledgments}


\pagebreak

\begin{widetext}

\begin{figure}[b]
\includegraphics[width=.6\textwidth]{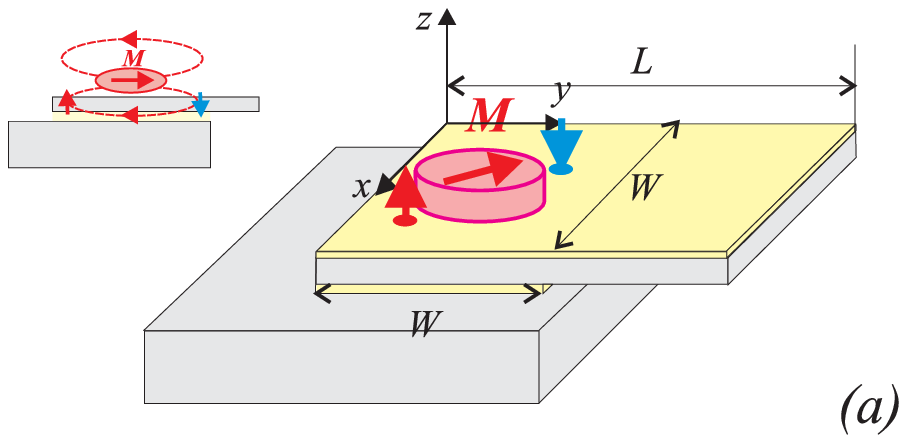}
\vspace{0.5cm}
\includegraphics[width=.6\textwidth]{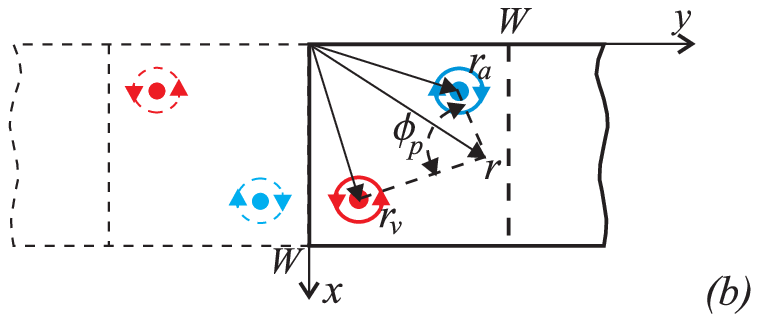}
\caption{(Color online)(a) Schematic diagram of the Josephson contact
with the ferromagnetic particle on the top electrode.
The junction area ($W\times W$) occupies only part of the
superconducting electrodes ($L \gg W$).
The vortices indicated by vertical arrows
are located near the opposite poles
of a uniformly magnetized particle.
The inset schematically shows the structure of the stray field
of the ferromagnetic particle and the pair of oppositely
directed Abrikosov vortices which pierce the
top superconducting electrode.
(b) The picture shows the location of a vortex ($\mathbf{r}_v$)
and antivortex ($\mathbf{r}_a$) trapped in the top electrode
of the contact.
The phase difference $\phi_p$ is specified by the angle between
the directions from the reference point $\mathbf{r}$ to the
points of the vortices location $\mathbf{r}_{v,a}$.
The thin dashed lines shows vortex(antivortex) images,
which provides zero normal currents at the edge $y=0$.}
\label{Fig:1}
\end{figure}
%
%
%
%
\begin{figure}
\includegraphics[width=0.35\textwidth]{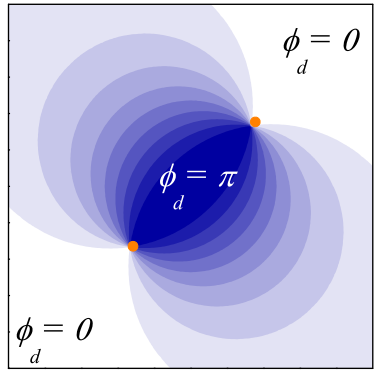}
\caption{(Color online) Distribution of cosine of the phase $\phi_p$
are schematically illustrated. The dark area corresponds to
the region where $\pi/2 < \phi_p < 3\pi/2$ and $\cos\phi_p < 0$.
The vortex(antivortex) position is shown by the bright spot.}
\label{Fig:2}
\end{figure}
%
%
%
%
\begin{figure}
\includegraphics[width=0.45\textwidth]{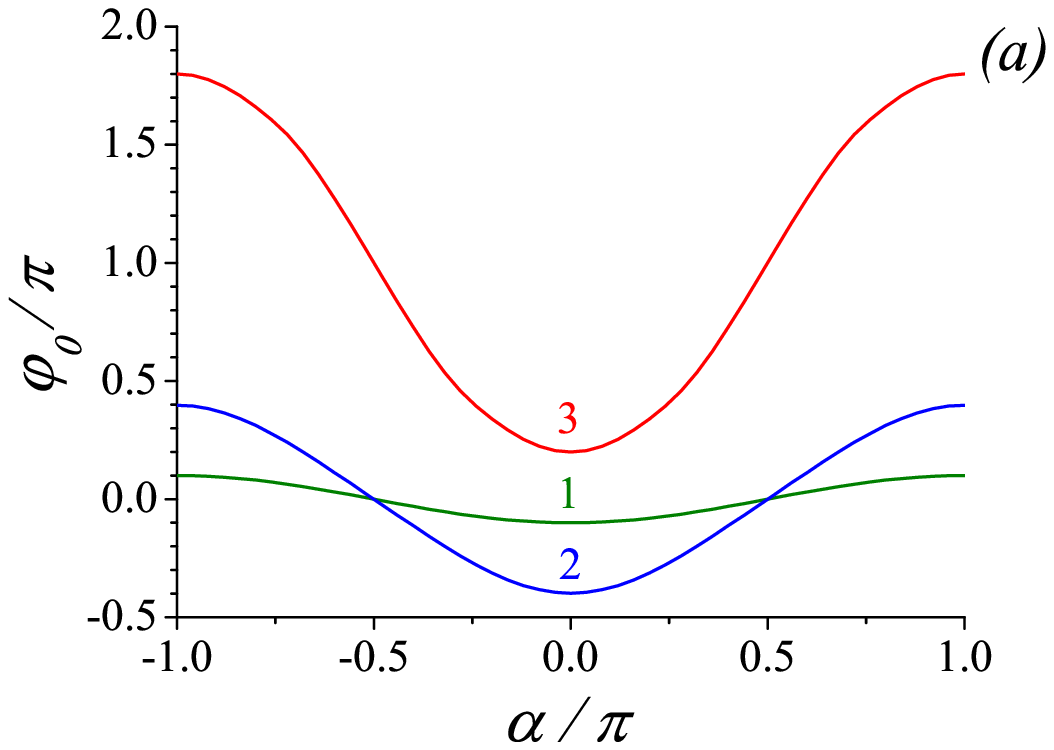}
\includegraphics[width=0.45\textwidth]{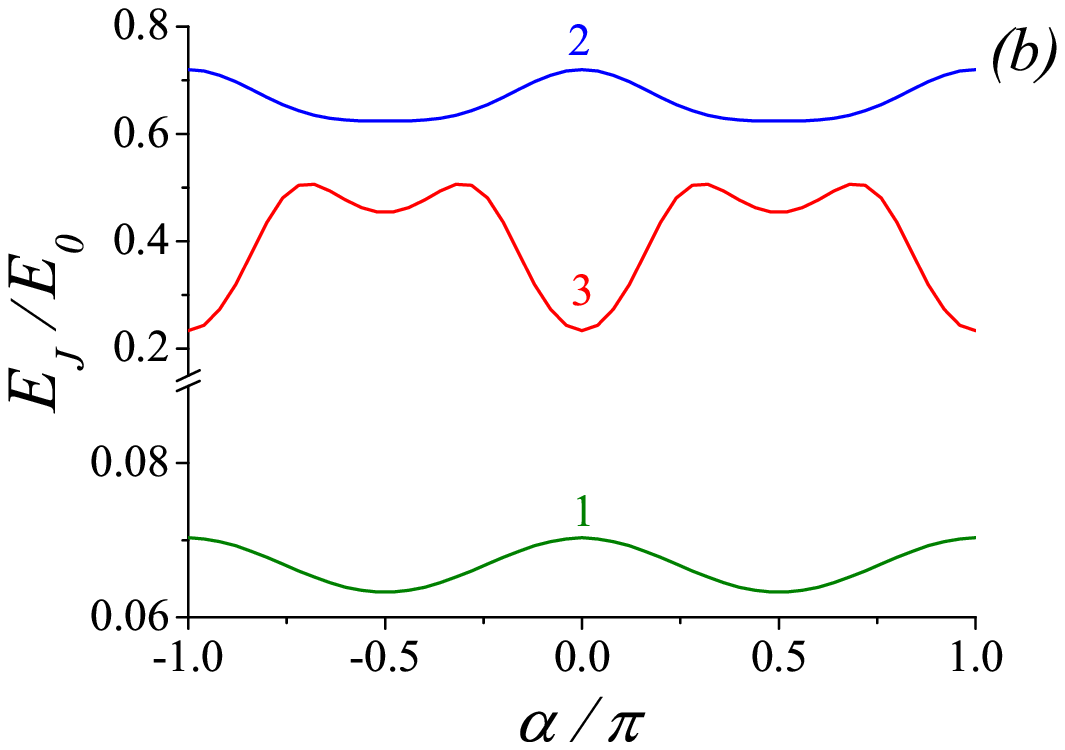}
\caption{(Color online) Dependence of the average phase difference
$\varphi_0$ (a) and the Josephson energy $E_J^{\varphi_0}$
of the ground state; (b) on the angle of the vortex--antivortex pair
rotation $\alpha$ for different values of the pair size $2a$:
1. $2a = 0.1 W$; 2. $2a = 0.4 W$; 3. $2a = 0.8 W$
($E_0=\hbar I_c / 2 e$).}
\label{Fig:3}
\end{figure}
%
%
%
\begin{figure}
\includegraphics[width=0.5\textwidth]{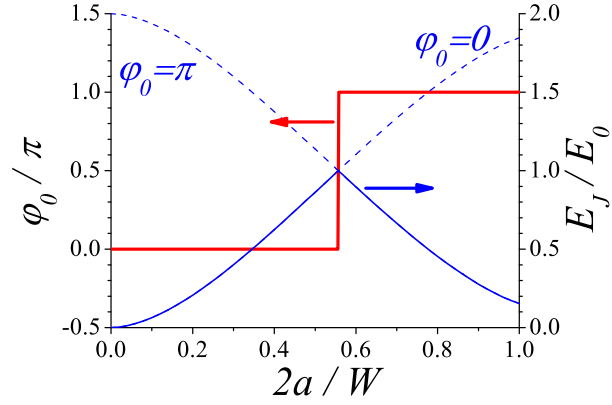}
\caption{Dependence of the average phase difference $\varphi_0$
and the Josephson energy $E_J^{\varphi_0}$ of the ground state
on the size $2a$ of the vortex-antivortex pair ($\alpha=\pi/2$).
The dashed lines show the dependence $E_J^0(2a)$ and $E_J^\pi(2a)$
for $\varphi_0=0$ and $\varphi_0=\pi$, respectively;
$E_J = \mathrm{min}\{E_J^0,\, E_J^{\pi}\}$
($E_0=\hbar I_c / 2 e$).}
\label{Fig:4}
\end{figure}
%
%
%
%
\begin{figure}
\includegraphics[width=0.338\textwidth]{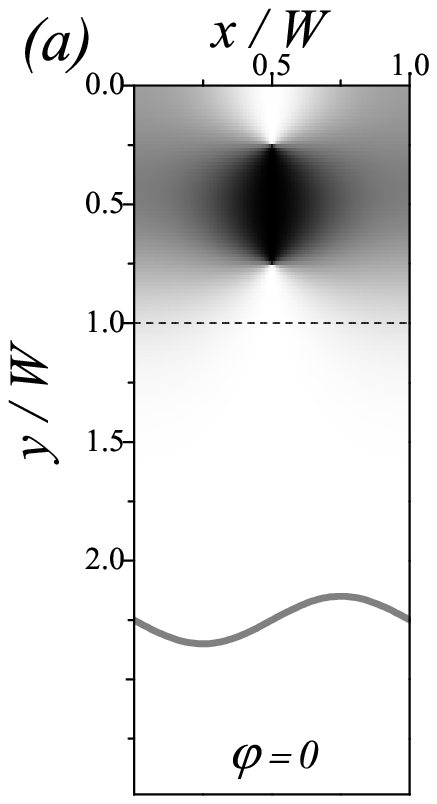}
\includegraphics[width=0.317\textwidth]{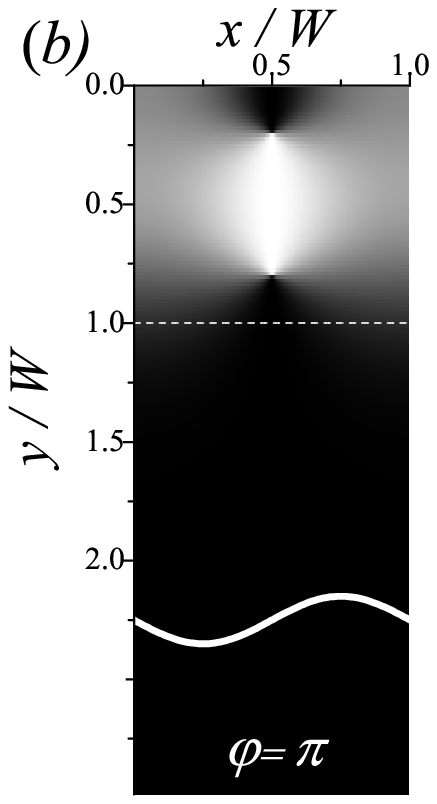}
\caption{Distribution of cosine of the phase
$\varphi=\phi(\mathbf{r})+\varphi_0$
for different values the pair size (intervortex distance)
$2a$: (a) $2a = \mathrm{0.5}\,W$ ; (b) $2a = \mathrm{0.6}\,W$.
The edge of the junction $y = W$ is marked by the dash line.}
\label{Fig:5}
\end{figure}
%
%

\end{widetext}

\end{document}